\documentclass[format=acmsmall, review=false, screen=true, nonacm=true]{acmart}
\usepackage{booktabs} 

\setcitestyle{acmnumeric}

\usepackage{xspace}
\usepackage{amssymb}
\usepackage{mathtools}
\usepackage[inline]{enumitem}

\usepackage[textwidth=12mm,obeyFinal,colorinlistoftodos]{todonotes}

\DeclareMathOperator{\poly}{{\rm poly}}

\DeclareMathOperator{\rank}{{\rm rank}}

\newcommand{\FPT}{{\sf FPT}\xspace}
\newcommand{\PA}{{\sf PA}\xspace}

\newcommand{\D}{{\sf D}\xspace}
\newcommand{\T}{{\sf T}\xspace}

\newcommand{\LL}{{\sf L}\xspace}
\newcommand{\NP}{{\sf NP}\xspace}

\def\ve#1{\mathchoice{\mbox{\boldmath$\displaystyle\bf#1$}}
{\mbox{\boldmath$\textstyle\bf#1$}}
{\mbox{\boldmath$\scriptstyle\bf#1$}}
{\mbox{\boldmath$\scriptscriptstyle\bf#1$}}}
\newcommand\vea{{\ve a}}

\newcommand\vecc{{\ve c}}

\newcommand\vem{{\ve m}}

\newcommand\ves{{\ve s}}
\newcommand\vet{{\ve t}}

\newcommand\vex{{\ve x}}
\newcommand\vey{{\ve y}}

\newcommand\veDelta{{\ve \Delta}}
\newcommand\vezero{{\ve 0}}

\def\Z{\mathbb{Z}}
\def\N{\mathbb{N}}

\def \R {{\mathbb{R}}}
\def \RR {\mathcal{R}}

\newcommand{\pref}{\ensuremath{\succ}}
\newcommand{\Oh}{\mathcal{O}}

\newcommand{\prob}[3]{
    \noindent
    \begin{center}
      \fbox{
        \begin{minipage}{.96\linewidth}
          \begin{tabularx}{\columnwidth}{lX}
	        \multicolumn{2}{l}{#1}\\
	        {\bf Input:}&{#2}\\
	        {\bf Find:}&{#3}
          \end{tabularx}
        \end{minipage}
      }
    \end{center}
}

\usepackage{nicefrac}
\usepackage{soul}
\usepackage{url}
\usepackage{tabularx}
\usepackage[utf8]{inputenc}
\usepackage{graphicx}
\usepackage{amsmath}
\usepackage{xcolor}
\usepackage{tasks}
\theoremstyle{definition}
\newtheorem{remark}{Remark}

\newcommand{\calR}{\mathcal{R}}

\newcommand{\lcm}{\textrm{lcm}}
\newcommand{\mypara}[1]{\smallskip\noindent\textbf{#1.}\quad}

\newcommand{\martin}[1]{\textcolor{blue}{Martin says: #1}}

\title{Multi-Party Campaigning}

\author{Martin Koutecký}
\affiliation{%
	\institution{Charles University}
	\city{Prague}
	\country{Czech Republic}}
\email{koutecky@iuuk.mff.cuni.cz}
\author{Nimrod Talmon}
\affiliation{%
	\institution{Ben-Gurion University}
	\city{Beer Sheva}
	\country{Israel}}
\email{talmonn@bgu.ac.il}
\renewcommand{\shortauthors}{M. Koutecký and N. Talmon}

\begin{abstract}
	We study a social choice setting of manipulation in elections and extend the usual model in two major ways: first, instead of considering a single manipulating agent, in our setting there are several, possibly competing ones; second, instead of evaluating an election after the first manipulative action, we allow several back-and-forth rounds to take place. We show that in certain situations, such as in elections with only a few candidates, optimal strategies for each of the manipulating agents can be computed efficiently.
	
	Our algorithmic results rely on formulating the problem of finding an optimal strategy as sentences of Presburger arithmetic that are short and only involve small coefficients, which we show is fixed-parameter tractable -- indeed, one of our contributions is a general result regarding fixed-parameter tractability of Presburger arithmetic that might be useful in other settings. Following our general theorem, we design quite general algorithms; in particular, we describe how to design efficient algorithms for various settings, including settings in which we model diffusion of opinions in a social network, complex budgeting schemes available to the manipulating agents, and various realistic restrictions on adversary actions.
\end{abstract}

\begin{document}

\maketitle

\section{Introduction}

Within computational social choice, the study of external agents wishing to rig a given election has received extensive study, most notably by studying problems of election control and bribery in elections~\cite{briberysurvey}:
  In these problems, an external agent aims at altering the result of a given election;
  in election control, such an agent can change the structure of the election, usually by adding/removing voters/candidates, while in bribery problems, such an agent can change the way certain voters vote.
  
Existing literature concentrates only on cases in which just one external agent (e.g., one briber) exists~\cite{faliszewski2009hard}. There are some papers which take a more game-theoretic approach, but from a cooperative perspective~\cite{bachrach2011coalitional}, while here we are concerned with a non-cooperative perspective.
Furthermore, all existing work (to the best of our knowledge) only deals with one action of bribery or control which is followed (immediately, or potentially with an additional opinion diffusion step~\cite{faliszewski2018opinion}) by evaluating the resulting election with respect to the agent's initial goal.

Here we are studying a very rich and realistic model, in which several competing agents, each agent with its own objective, alter the election in multiple rounds, responding to each other's actions.
In particular, our model captures the setting of multi-party campaigning, in which several parties are campaigning over the same election. 
In the most basic model we consider $k$ bribers operating on an election with $m$ alternatives and $n$ voters. There is a process, running for $\ell$ turns, in which, in each turn, each briber can change certain votes of their choice, respecting some budget limit. At the end of the process, a winner of the resulting election is chosen according to some voting rule $\calR$. Our goal is to compute optimal bribery strategies for each agent.

To this end, we construct a Presburger arithmetic formula (for background cf. a recent guide~\cite{haase2018survival}) that is true if and only if a given agent has a winning strategy which respects the budget; indeed, there is a procedure to recover this strategy if it exists.
Furthermore, by ensuring that this formula is short and contains only small coefficients, we are able to compute such optimal strategies efficiently;
this follows by combining a careful analysis of the algorithm of Cooper~\cite{cooper1972theorem} for deciding Presburger arithmetic formulas together with algorithms for convex integer optimization in small dimension~\cite{GLS,DadushPV:2011}.
Our contribution here is a complexity analysis of deciding Presburger arithmetic formulas, taking into account multiple parameters, showing that only some of them undergo a (necessary) combinatorial explosion.
In particular, we infer that deciding Presburger arithmetic and even optimizing a convex function over its satisfying assignments is fixed-parameter tractable parameterized the length of the given formula and its largest coefficient.
A problem $\Pi$ is \emph{fixed-parameter tractable (\FPT) parameterized by $k$} if it has an algorithm running in time $f(k) \poly(n)$ for any instance of size $n$.
Classifying a problem as \FPT gives a formal way of saying that a special class of instances (those with small values of the parameter $k$) are easy to solve for a problem which is hard in general.

In essence, Presburger arithmetic contains logical formulas over linear constraints (this stands in contrast with Peano arithmetic, which also permits multiplication of variables).
Thus, it is possible to design the formulas described above in a modular way incorporating ``hooks'' within the basic model described above, which allow future extensions by plugging in more complex formulas; we later demonstrate several such extensions using this framework.
In particular, we show that our basic model can be drastically extended to include settings in which \begin{enumerate*}[label=\bfseries(\arabic*)] \item voter types represent complex voter differences; \item the budgets available to the agents change between turns, perhaps depending on the intermediate election at each turn, representing campaign polling and fundraising interaction; \item settings in which voters are embedded in a social network and a diffusion process causes voters to update their votes; and \item settings in which various restrictions on the strategies available to the agent are present, allowing for more realistic modeling.
\end{enumerate*}
Importantly, for these extensions, computational efficiency follows similarly to the basic model.

\mypara{Our Contributions}
Our main contributions are as follows:
\begin{enumerate*}[label=\bfseries(\arabic*)]
  \item We suggest a useful model that captures the natural setting of multi-party campaigning, including rich model settings with,
  e.g., opinion diffusion in social networks and involved budgeting schemes;
  \item We devise efficient algorithms for finding optimal strategies for agents in these models, by a reduction to optimization over satisfying assignments of formulas in Presburger arithmetic;
  \item We show that Presburger arithmetic convex minimization is fixed-parameter tractable wrt.\ the length of the given formula and the absolute value of its largest coefficient.
\end{enumerate*}

\subsection{Related Work}

There is a vast literature on control and bribery~\cite{briberysurvey}:
  In particular, Elkind et al.~\cite{elkind2009swap} proposed the notion of swap bribery, in which the external agent (i.e., the briber) pays for each swap it causes in a voter's preference order. Faliszewski et al.~\cite{faliszewski2009hard} further studied the complexity of various bribery problems. 
To the best of our knowledge, there are no works dealing with multiple external agents manipulating a given election.
(We do mention recent work that consider multiple bribers~\cite{zhou2019briber,grandi2016network,grandi2018complexity}, however in the different setting of a ranking system; in particular, our technical contribution differs largely from these works.)
  
Our algorithmic techniques allow us to show fixed-parameter tractability (\FPT) of our problems. \FPT algorithms for bribery problems have been studied quite extensively~\cite{bredereck2016complexity,dorn2012multivariate}, also for more complex social choice settings such as multiwinner elections~\cite{faliszewski2017bribery}. We differ greatly from these works as we consider the more general case containing several manipulating agents and multiple back-and-forth rounds.
Related, as we show later, our model can accommodate certain diffusion processes that occur, e.g., when agents live in a social network.
In this context, we mention the work of Bredereck and Elkind~\cite{bredereck2017manipulating} and Faliszewski et al.~\cite{faliszewski2018opinion} on the interchange between bribery actions and opinion diffusion in social networks.

Our algorithmic framework is presented for the representation of elections as society vectors. The concept of a society vector was used by Knop et al.~\cite{knop2018unifying} for studying bribery problems and was also used later by Faliszewski et al.~\cite{faliszewski2018opinion}. This representation is very general and serves as a unifying framework to describe extensions to our basic model

Algorithmically, our approach is based on formulating combinatorial problems as (linear) minimization over satisfying assignments of Presburger arithmetic formulas.
In this context, the overall idea is quite similar to the use of Lenstra's celebrated result~\cite{lenstra} regarding fixed-parameter tractability of deciding Integer Linear Programming parameterized by the number of integer variables.
Other fixed-parameter tractable algorithms from the theory of integer programming which have found use in computational social choice are algorithms for $n$-fold IP~\cite{EisenbrandEtAl2019} and parametric ILP~\cite{eisenbrand2008parametric}.
Some recent work in computational social choice that uses some of these tools and share the same general flavor~\cite{bredereck2020mixed,knop2018voting,knop2017combinatorial}.

Presburger Arithmetic has been introduced by Presburger and Tarski in 1929~\cite{presburger1929uber} and we built on Cooper's algorithm from 1972~\cite{cooper1972theorem}.
Nguyen and Pak~\cite{nguyen2019short} recently resolved a major open problem by proving that even deciding short PA formulas is \NP-hard (and beyond) if they are allowed to contain large coefficients.
Klaedtke~\cite{klaedtke2008bounds} gave more refined bounds for the automata approach to deciding PA formulas, however, even his bounds do not make the distinction between coefficients and constants which is necessary to obtain our algorithmic result.

\section{Presburger Arithmetic is Fixed-Parameter Tractable}

Let $m,n$ be integers.
We define \mbox{$[m,n] := \{m, m+1, \ldots, n\}$} and \mbox{$[n] := [1,n]$}.
Throughout, we reserve bold face letters (e.g., $\vex,\vey$) for vectors.
For a~vector $\vex$ its $i$-th coordinate is $x_i$.
We write $\vea \vex$ for the dot product of vectors $\vea$ and $\vex$.

We wish to develop efficient algorithms that find optimal strategies for agents that are manipulating a given election. 
Our approach is to write a formula in Presburger arithmetic (here we shorten to PA; this is not to be confused with Peano arithmetic) with a vector $\vem$ of free variables such that the satisfying assignments are bribery actions corresponding to a first move in a winning strategy.
Here we first provide a brief introduction to PA, and later show that optimizing over the satisfying assignments of a PA formula can be done efficiently if some of its parameters are bounded, as will be the case for the formulas modeling winning strategies.

PA is a useful logic to reason about numbers. Intuitively, PA can be viewed as Integer Linear Programming (ILP) enriched with logical connectives and quantifiers.
For two formulas $\Phi$ and $\Psi$, we denote their equivalence by $\Phi \cong \Psi$.

\begin{definition}[Extended Presburger Arithmetic (PA)]
An \emph{atom} (or \emph{atomic formula}) is a linear inequality $\vea \vex \leq b$ or a congruence $\vea \vex \equiv b \mod p$, with $\vea \in \Z^n$ and $b,p \in \Z$.
We call $\vet \equiv \vea \vex + b$ for some $\vea$ and $b$ a \emph{term}.
A \emph{formula} is obtained by taking Boolean combinations of atoms using the standard logical connectives ($\wedge, \vee, \implies, \neg$, etc.) and by existential and general quantifiers $\exists, \forall$, respectively.
Denote by $\PA$ the set of all PA formulas.
A \emph{literal} is an atom or its negation.
A variable is \emph{bound} in a formula $\varphi$ if it appears in a quantifier, and it is \emph{free} otherwise.
If $\vex$ is a vector of the free variables of a formula $\varphi$, we write $\varphi(\vex)$.
\end{definition}

\begin{remark}
The term ``extended'' in the definition above refers to the congruence atoms that are not present in the original language as defined by Presburger~\cite{presburger1929uber}; however,
it is typical to speak of PA as this extended language because it allows for quantifier elimination, unlike PA without congruence atoms.
For detailed definitions cf. Klaedtke~\cite{klaedtke2008bounds}.
\end{remark}

We provide some further useful notation below.
For $\varphi \in \PA$ we define $\T(\varphi)$ to be the set of all atoms of $\varphi$ of the form $\vea \vex \leq b$, $\D(\varphi)$ to be the set of all atoms of the form $\vea \vex \equiv b \mod p$, $\LL(\varphi)$ to be the number of symbols of $\varphi$ (i.e., the number of atoms, logical connectives, and quantifiers),\footnote{Note that this definition is different than the standard definition of the length of a formula, which uses unary encoding of numbers.} the \emph{maximum coefficient} $\alpha(\varphi)$ to be the maximum $\|\vea\|_\infty$ and $p$ contained in any of its atoms, and the \emph{maximum constant} $\beta(\varphi)$ to be the largest right hand side $b$ in any of its atoms.

The problem we wish to solve is the following:
\prob{\textsc{Presburger arithmetic minimization}}{A PA formula $\varphi(\vex)$ with $d$ free variables, and a function $f: \R^d \to \R$.}
{Find an assignment $\vex \in \Z^d$ satisfying $\varphi(\vex)$ and minimizing $f(\vex)$ (among satisfying assignments).}

The main algorithmic result regarding Presburger arithmetic we prove here is the following.

\begin{theorem}\label{theorem:pafpt}
  \textsc{Presburger arithmetic minimization} is fixed-parameter tractable parameterized by $\LL(\varphi) + \alpha(\varphi)$ for any convex function $f$.
\end{theorem}
\begin{remark}
The proof combines two elements.
First, we perform a careful analysis of Cooper's algorithm~\cite{cooper1972theorem} which uses quantifier elimination to find a quantifier-free formula $\psi$ equivalent to $\varphi$.
We show that $\LL(\psi), \alpha(\psi) \leq f(\LL(\varphi), \alpha(\varphi))$ and $\beta(\psi) \leq f(\LL(\varphi), \alpha(\varphi)) \cdot \beta(\varphi)$.
While Cooper's algorithm is by now textbook material~\cite{bradley2007calculus}, it has not yet been recognized as a fixed-parameter algorithm, and the complexity bound we provide is not explicitly stated (as far as we know) in any existing work.
Second, $\psi$ is transformed into disjunctive normal form (DNF), yielding a formula $\psi_1 \vee \psi_2 \vee \cdots \vee \psi_K$ such that an assignment $\vex$ satisfies $\psi$ if and only if it satisfies some $\psi_i$, $i \in [K]$, with each $\psi_i$ being a conjunction of linear inequalities or congruences.
Such a conjunction can be then turned into a system of only linear constraints (by linearizing the congruences) and then one can apply any \FPT algorithm for convex minimization in small dimension~\cite{GLS,DadushPV:2011} to each of these systems and return the best result among all.
\end{remark}
\begin{proof}[{Proof of Theorem~\ref{theorem:pafpt}}]
Assume that $\varphi(\vey) \equiv Q_1 x_1 Q_2 x_2 \dots Q_{k-1} x_{k-1} Q_k x_k \zeta(\vex,\vey)$, with $Q_1, \dots, Q_k \in \{\exists, \forall\}$ and $\zeta(\vex,\vey)$ containing no quantifiers. Here, $\vey$ is the vector of free variables of $\varphi$.
The proof proceeds by \emph{quantifier elimination}: if we show that the innermost quantifier ($\exists x_k$ in our example) can be eliminated, i.e., if we can construct an equivalent formula $\varphi'(\vey) \equiv Q_1 x_1 \dots Q_{k-1} x_{k-1} \zeta'(x_1, \dots, x_{k-1}, \vey)$, then repeatedly applying this procedure reduces $\varphi$ down to a formula with no quantifiers and only containing variables $\vey$ (but no variables $\vex$).
Optimizing over the satisfying assignments of such a formula then (with some more work) reduces to optimization over linear constraints in small dimension.
Already the original proof of Presburger that \PA is decidable worked by quantifier elimination.
We shall now describe an algorithm of Cooper, which achieves better complexity.
Let us stress that our goal is not to prove the correctness of the algorithm, only to describe it in sufficient detail so as to analyze its complexity, and still convey the main underlying intuition; for correctness, we refer the reader to existing textbooks (e.g.,~\cite{bradley2007calculus}).

Consider a formula $\varphi(x_k) \equiv \exists x_k \zeta(x_k)$, where $\zeta(x_k)$ is quantifier-free.
Here, $\varphi$ stands for the suffix of the whole formula to be decided; we sometimes disregard the prefix $Q_1 x_1 \cdots Q_{k-1} x_{k-1}$ and the free variables $\vey$ for brevity.
Note that if the last quantifier was $\forall$, we negate the formula, obtain $\exists$ as the last quantifier, and in the end negate the (eventually quantifier-free) formula again.
The algorithm proceeds in three steps.
\textbf{First}, we put $\varphi(x_k)$ into negation normal form (pushing all negations inward as much as possible) using De Morgan's rules, yielding an equivalent formula $\varphi_1(x_k)$.
\textbf{Second}, we normalize $\varphi_1(x_k)$ so that the coefficients of $x_k$ are all $1$ or $-1$, yielding $\varphi_2(x_k)$.
This is done as follows.
Let $A$ be the set of coefficients of $x_k$ in $\varphi_1(x_k)$, and let $M = \lcm(A)$, where $\lcm$ is the least common multiple and hence $M \leq (\max_{a \in A} a)^{|A|}$.
Replace every atom $\vea (\vex,\vey) \leq b$ in $\T(\varphi_1)$ with $(M/a_k) \cdot \vea (\vex,\vey) \leq (M/a_k) b$ (recall $a_k$ is the coefficient of $x_k$ in $\vea \vex$), replace every atom $\vea (\vex,\vey) \equiv b \mod p$ in $\D(\varphi_1)$ with $(M/a_k) \vea (\vex,\vey) \equiv (M/a_k)b \mod (M/a_k)p$ and call $\varphi_2(x_k)$ the resulting formula.
Now we perform the substitution $x'_k = M x_k$, hence let $\varphi_3(x'_k) \equiv (\varphi_2(Mx_k) \wedge x'_k \equiv 0 \mod M)$.
Now, all coefficients of $x'_k$ in $\varphi_3(x'_k)$ are $1$ or $-1$.

The \textbf{third} step is the most involved.
Denote by $\bar{\vex} = (x_1, \dots, x_{k-1})$.
Notice that all literals of $\varphi_3(x'_k)$ are (perhaps after simple rearranging) of one of the following types, where $\vet_1, \vet_2, \vet_3$ are terms over the variables $\bar{\vex}$:
\settasks{counter-format = (tsk),}
\begin{tasks}(2)
\task $x'_k \leq \vet_1$,\label{terms:A}
\task $\vet_2 \leq x'_k$,\label{terms:B}
\task $x'_k = \vet_3 \mod p$,\label{terms:C}
\task $\neg(x'_k = \vet_3 \mod p)$.\label{terms:D}
\end{tasks}

We distinguish two cases. Either $\varphi_3(x'_k)$ has arbitrarily small satisfying assignments (i.e., for any $t \in \Z$, there exists a satisfying assignment to $x'_k$ smaller than $t$).
Then, for a sufficiently small satisfying assignment, literals of type~\ref{terms:A} are satisfied and can be replaced by $\top$, and literals of type~\ref{terms:B}  are falsified and can be replaced by $\bot$.
Call $\varphi_{-\infty}(x'_k)$ a formula obtained from $\varphi_3(x'_k)$ by the aforementioned replacements.
Let $M'$ be the least common multiple of all the moduli $p$ in literals of types~\ref{terms:C} and~\ref{terms:D}, and let $\varphi_{41} \equiv \bigvee_{j=1}^{M'} \varphi_{-\infty}(j)$.
Then $\varphi_{41}$ is satisfiable iff $\varphi(x_k)$ has arbitrarily small satisfying assignments.

Now we construct a formula $\varphi_{42}$ which is satisfied in the converse case when $\varphi(x_k)$ has a least satisfying assignment.
For such an assignment some type-\ref{terms:B} literal is satisfied and for smaller assignments it is not.
We let $B = \{\vet_2 \mid \vet_2 \leq x'_k \text{ is a type-\ref{terms:B} literal}\}$ and define $\varphi_{42} \equiv \bigvee_{j=1}^{M'} \bigvee_{\vet_2 \in B} \varphi_3(\vet_2 + j)$ (note here that $\varphi_3(\vet_2 + j)$ is $\varphi_3$ with $x'_k$ substituted by $\vet_2 + j$).
Then $\varphi_4 \equiv \varphi_{41} \vee \varphi_{42}$ and it does not contain any occurrence of $x_k$, finishing the elimination.

Let us now bound the length, maximum coefficient, and maximum constant of the formula resulting after eliminating all quantifiers.
For this, it suffices to bound the blow-up caused by one quantifier elimination.
Step one does not consider the coefficients or constants in any way and hence $\LL(\varphi_1)$ is  bounded by a function of $\LL(\varphi)$, $\alpha(\varphi_1) = \alpha(\varphi)$, and $\beta(\varphi_1) = \beta(\varphi)$.
Step two increases the coefficients and constants by a number only depending on the previously largest coefficient and the number of literals and adds two symbols, specifically, $\alpha(\varphi_3) \leq \alpha(\varphi_1)^{\LL(\varphi_1)}$, $\beta(\varphi_3) \leq \beta(\varphi_1) \cdot \alpha(\varphi_1)^{\LL(\varphi_1)}$, and $\LL(\varphi_3) = \LL(\varphi_1) + 2$.
In step three again the length, largest coefficient and largest constant only grows by a factor of the initial length and largest coefficient, specifically, $\LL(\varphi_4) \leq \alpha(\varphi_3)^{\LL(\varphi_3)} \cdot \LL(\varphi_3)^2$, $\alpha(\varphi_4) \leq \alpha(\varphi_3)^{\LL(\varphi_3)} \cdot \alpha(\varphi_3) = \alpha(\varphi_3)^{\LL(\varphi_3)+1}$, and $\beta(\varphi_4) \leq \beta(\varphi_3) \cdot \alpha(\varphi_3)^{\LL(\varphi_3)}$.
Applying the bounds derived above inductively, we get the following intermediate claim:
\begin{lemma}
There exists a computable function $g$ such that given a formula $\varphi(\vey) \in \PA$, an equivalent quantifier-free formula $\psi(\vey)$ can be obtained in time $\Oh(g(\LL(\varphi), \alpha(\varphi)))$ and it satisfies:
\begin{itemize}
\item $\LL(\psi), \alpha(\psi) \leq g(\LL(\varphi), \alpha(\varphi))$,
\item $\beta(\psi) \leq g(\LL(\varphi), \alpha(\varphi)) \cdot \beta(\varphi)$.
\end{itemize} 
\end{lemma}

Now, we come to the second step, where we wish to optimize over the satisfying assignments of $\psi(\vey)$.
We transform $\psi$ to disjunctive normal form (so that it is a disjunction of conjunctions of atoms), which may increase its length exponentially, but that is still bounded by a function of the parameters.
Let us now assume that $\psi \equiv \psi_1 \vee \psi_2 \cdots \psi_K$ for some $K \in \N$, where each $\psi_i$ is a conjunction of literals.
Clearly, an assignment $\vey$ minimizing $f(\vey)$ over the satisfying assignments of $\psi$ satisfies some conjunction $\psi_i$, so we may instead minimize separately over the satisfying assignments for each $\psi_i$, $i \in [K]$.
This can be done by algorithms for convex integer minimization whenever $\psi_i$ is a conjunction of linear atoms~\cite{GLS,DadushPV:2011}, so our next task is to linearize the congruence atoms.

This is easy for positive literals: say we have $\vea \vey = b \mod p$; then we introduce a new variable $z$, and add a linear constraint $pz = \vea \vey - b$, which is satisfied iff $\vea \vey - b$ is divisible by $p$.
For the negative literals, this is a little bit trickier: say we have $\vea \vey \neq b \mod p$.
This is equivalent to saying that $\vea \vey \mod p$ is between $b+1$ and $b+p-1$.
We introduce two variables $z, z'$ and add the following three linear constraints: $pz = \vea \vey - z'$ and $b+1 \leq z' < b+p$.
Since we have introduced a constant number of new variables and constraints for each congruence atom, the system resulting from $\psi_i$ is still of length bounded by a function of the parameters, and evaluating $f$ over its integer assignments can be done in fixed-parameter tractable time (specifically, in time $\delta^{\Oh(\delta)} \poly\log (\alpha(\psi_i), \beta(\psi_i))$ where $\delta$ is the number of variables, which is at most $3d$, the dimension of $\vey$, i.e., the number of free variables of $\varphi$)~\cite{GLS,DadushPV:2011}.
This concludes the proof.
\end{proof}

\section{A Basic Model with Multiple Bribers}\label{section:basic}

Here we describe a basic model of an election with multiple manipulating agents. Informally, we wish to model a situation containing a set of voters where each manipulating agent wishes to make its preferred candidate win in an election that occurs eventually, and, to this end, can, at a certain cost, alter the opinions of several voters.
We first provide some preliminaries and then describe our formal model.

\subsection{Elections, Bribers, and Societies}\label{sec:preliminaries}
\mypara{Elections}
An (ordinal) election~$(C,V)$ consists of a set $C$ of candidates and a set~$V$ of voters, who indicate their preferences over the candidates in~$C$, represented via a \emph{preference order} $\pref_v$ which is a total order over~$C$.
We often identify a voter $v$ with her preference order~$\pref_v$.
Denote by $\textrm{rank}(c,v)$ the rank of candidate~$c$ in~$\pref_v$; $v$'s most preferred candidate has rank 1 and her least preferred candidate has rank $|C|$.
For distinct candidates~$c,c'\in C$, write $c\pref_v c'$ if voter~$v$ prefers~$c$ over~$c'$ (i.e., $v$ ranks $c$ higher than she ranks $c'$).

\mypara{Voting rules}
A voting rule~$\RR$ is a function that maps an election $(C,V)$ to a subset $W\subseteq C$, called the \emph{winners}. Many voting rules have been considered in the social choice literature.
Perhaps the simplest voting rule is Plurality, where the winner is an alternative which is ranked first by the largest number of voters.
As another example, the \emph{Borda rule} selects as a winner a candidate whose average ranking over the voters is the highest, that is, a candidate $c$ gets $m-\rank(c,v)$ from each candidate $v$, and the candidate with the most points wins.

\mypara{Swaps and Swap Bribery}
Let $(C,V)$ be an election and let $\pref_v\in V$ be a voter.
For candidates $c,c'\in C$, a \emph{swap} $s = (c,c')_v$ corresponds to an exchange between the positions of $c$ and $c'$ in~$\pref_v$; denote the perturbed order by $\pref_v^s$.
A swap~$(c,c')_v$ is \emph{admissible in $\pref_v$} if $\rank(c,v) = \rank(c',v)-1$ (note that this is indeed a swap of consecutive candidates).
A set $S$ of swaps is \emph{admissible in $\pref_v$} if each swap in $S$ can be applied sequentially in~$\pref_v$, one after the other, in some order, such that each one of them is admissible.
Note that the perturbed vote, denoted by $\pref_v^S$, is independent from the order in which the swaps of $S$ are applied.
We extend this notation for applying swaps in several votes and denote it $V^S$ (note that, in particular, the cost of $V^S$ equals the sum of costs of $s \in S$).
We specify $v$'s cost of swaps by a function $\sigma^v: C\times C\rightarrow \mathbb{Z}$.

In the Swap Bribery problem, which we generalize here, we are given an election $(C, V)$, a designated candidate $c^{\star} \in C$, and swap costs $\sigma^v: C\times C\rightarrow \mathbb{Z}$ for $v \in V$. The goal is to identify a set $S$ of admissible swaps of minimum cost so that $c^{\star}$ wins the election $(C, V^S)$ under the rule~$\RR$.

\mypara{Societies and Moves}
It will be useful to view an electorate not simply as a set of votes, but bundled by voter types.
Let $\tau \in \N$ be the number of \emph{types of voters}; note that $\tau \leq n$, and it can be significantly smaller. E.g., if voters are only distinguished by their preference orders, then $\tau \leq m!$ which might be much smaller than $n$.

A \emph{society} is a non-negative $\tau$-dimensional integer vector $\ves = (s_1, \dots, s_\tau)$, where $s_j$, $j \in [\tau]$, corresponds to the number of voters of type $j$ in the society.
In most problems, we are interested in modifying a society by moving people among types.
A \emph{move} is a vector $\vem = (m_{1,1}, \dots, m_{\tau, \tau})\in\mathbb Z^{\tau^2}$.
Intuitively, $m_{i,j}$ is the number of people of type $i$ turning type~$j$.

\begin{definition}
  A \emph{change} is a vector $\veDelta = (\Delta_1, \dots, \Delta_\tau)\in\mathbb Z^{\tau}$ whose elements sum up to $0$.
  We say that \emph{$\veDelta$ is the change associated with a move $\vem$} if, for all $i \in [\tau]$, $\Delta_i = \sum_{j=1}^\tau m_{j,i} - m_{i,j}$, and we write $\veDelta = \Delta(\vem)$.
  A change $\veDelta$ is \emph{feasible with respect to society $\ves$} if $\ves + \veDelta \geq \mathbf{0}$, i.e., if applying the change $\veDelta$ to $\ves$ results in a society (in other words, as long as there are enough voters from each type to move to other types).
\end{definition}

Another useful notion is the \emph{move costs vector}, which is a vector $\vecc = (c_{1,1}, \dots, c_{\tau, \tau})$ in $(\N \cup \{+\infty\})^{\tau^2}$ that satisfies the triangle inequality, i.e., $c_{i,k} \leq c_{i,j} + c_{j,k}$ for all distinct $i,j,k$.

\begin{remark}
In this work we focus on moves which correspond to swap bribery actions.
However, the bribery, manipulation, and control actions expressible as moves in societies are much more general, as shown by Knop et al.~\cite{knop2018unifying}.
For example, one may create, for each voter type $t \in [\tau]$, an ``inactive'' variant $t'$, and moving a voter from $t$ to $t'$ corresponds to deleting this voter while moving a voter from $t'$ to $t$ corresponds to adding it -- which are the actions considered in the CCAV and CCDV problems (constructive control by adding/deleting voters).
Hence, we encourage the reader to keep in mind that whenever we talk about swaps and bribery, many other types of actions may be substituted or added in that place.
\end{remark}

\subsection{A Formal Model}

In our formal model we consider a society with $n$ voters over $m$ candidates. Bundling voters into types according to their preferences, we have that the number of voter types equals the number of preference orders existing in the society (thus, in particular, upper bounded by $m!$). We consider $k$ bribers, $A_1, \ldots, A_k$, and a process of $\ell$ turns, such that, in each turn, each briber is given a budget of $B$ to be used to change the society; in each turn, the bribers bribe the society in a round-robin fashion -- $A_1$ first, then $A_2$, and so on.
The bribery operations are all unit-cost swap bribery operations;
in our context this means that, in each turn, each briber can cause at most $B$ swaps of consecutive candidates.
At the end of the $\ell$-th round, we apply the Borda voting rule on the society to identify a winning candidate.

We are interested in an optimal strategy for $A_1$; we refer to $A_1$ as \emph{our briber} where, w.l.o.g., the preferred candidate of $A_1$ is $p$.

\begin{remark}
It is also possible to model the situation when a different agent plays the first turn, i.e., where they play and we respond, and so forth.
In that case, we are able to verify, with the same complexity as all the results below, whether a winning strategy of our briber exists within a certain cost, but since the move that should be made will depend on the previous moves of the other bribers, there is no answer to be output other than ``yes/no''.
After a move by the other bribers is realized, we are in the original setting, because it is our turn, and we can compute the optimal response.
\end{remark}

\section{Optimal Strategies via Presburger Arithmetic}

Here we wish to formulate the problem of finding optimal strategies for our briber in the model described above using formulas in PA; this will be useful algorithmically, as Theorem~\ref{theorem:pafpt} then implies efficient algorithms if the relevant parameters are kept small.

First, we note that here we are interested in a worst-case adversarial model, so we assume that the only aim of the other bribers (i.e., $A_2, \ldots, A_k$) is to interfere with our briber; thus, while $A_1$ uses his bribery budget to try to make $p$ win the election under $\calR$ after the $\ell$ turns, all other bribers use their bribery budget to prevent alternative $p$ from winning.
Thus, in particular, the other bribers can collude and join forces (possibly also transferring money between them) to prevent $p$ from winning. So, in fact, the number of other bribers does not make a difference in our current worst-case model; in particular, $k - 1$ other bribers, each with a budget of $B$ for each turn, are equal, from the point of view of our briber, to a single briber with a budget of $(k - 1)B$ for each turn.
Thus, below we assume that there is only one other briber, which we refer to as \emph{the other briber}.

\subsection{Formulating Optimal Strategies}

Below we describe a formula $\Phi$ in PA; later we argue that $\Phi$ is satisfiable iff there is a strategy for our briber guaranteeing that our candidate wins.

We first describe some ingredients of $\Phi$.
First, $\ves_1^1$ is the initial society. Then, $\vecc$ is the cost vector corresponding to unit-cost swap-briberies (i.e., $c_{i,j}$ is the cost of swapping from type $i$ to type $j$, which is the number of inversions between the corresponding two permutations).
We need the following auxiliary predicates:
\begin{itemize}
\item $\textrm{PossibleMove}(\vem, \ves)$ is true for a move $\vem$ and a society $\ves$ if the resulting vector is a society, that is, $\textrm{PossibleMove}(\vem, \ves) \equiv \ves + \veDelta(\vem) \geq \vezero$.
\item $\textrm{FeasibleMove}(\vem, B)$ is true for a move $\vem$ and an integer $B$ if the number of swaps in move $\vem$ is at most $B$ that is, $\textrm{FeasibleMove}(\vem, B) \equiv \vecc \vem \leq B$.
\item $\textrm{ApplyMove}(\vem, \ves, \ves')$ is true for a move $\vem$ and two societies $\ves, \ves'$ if $\ves'$ is the result of applying $\vem$ to $\ves$, that is, $\textrm{ApplyMove}(\vem, \ves, \ves') \equiv \ves' = \ves + \veDelta(\vem)$. Since we prefer to view this as a function, we write $\ves' = \textrm{ApplyMove}(\vem, \ves)$.
\item $\textrm{BordaWinner}(p, \ves)$ is true for a society $\ves$ if $p$ is the Borda winner in it. This is encoded by observing that the Borda score of candidate $c$ is $S_c = \sum_{t \in [\tau]} \left(m-\rank(c, t)\right)$ (where $\rank(c,t)$ is the rank of $c$ for voters of type $t$), and $p$ is the unique Borda winner if $S_p > S_c$ for every candidate $c \neq p$, i.e., $\textrm{BordaWinner}(p, \ves) \equiv \bigwedge_{c \in C \setminus \{p\}} S_c < S_p$. Note that $S_c$ is just a shorthand for the aforementioned sum, so we need not introduce any new variables.
\end{itemize}
We are ready to describe the general structure of $\Phi$:
\begin{align*}
  \Phi(\vem_1^1) &\equiv \textrm{PossibleMove}(\vem_1^1, \ves_1^1) 
  \land \textrm{FeasibleMove}(\vem_1^1, B) 
  \land \ves_1^2 = \textrm{ApplyMove}(\vem_1^1, \ves_1^1) \\
  \forall \vem_1^2 &: \textrm{PossibleMove}(\vem_1^2, \ves_1^2)
  \land \textrm{FeasibleMove}(\vem_1^2, B) 
  \land \ves_2^1 = \textrm{ApplyMove}(\vem_1^2, \ves_1^2) \\
  &\vdots \\
  \exists \vem_\ell^1 &: \textrm{PossibleMove}(\vem_\ell^1, \ves_\ell^1) 
  \land \textrm{FeasibleMove}(\vem_\ell^1, B) 
  \land \ves_\ell^2 = \textrm{ApplyMove}(\vem_\ell^1, \ves_\ell^1) \\
  \forall \vem_\ell^2 &: \textrm{PossibleMove}(\vem_\ell^2, \ves_\ell^2)
  \land \textrm{FeasibleMove}(\vem_\ell^2, B) 
  \land \ves' = \textrm{ApplyMove}(\vem_\ell^2, \ves_\ell^2) \\
  &\land \textrm{BordaWinner}(p, \ves')   
\end{align*}

In $\Phi$, $\vem_i^j$ represents the $i$-th move of our briber, if $j = 1$, or the other briber, if $j = 2$. $\ves_i^j$ represents the society just before the $i$-th turn of our briber, if $j = 1$, or the other briber, if $j = 2$. Finally, $s'$ represents the eventual society.

Note that the lines beginning with ``$\exists$'' are for our briber while the lines beginning with ``$\forall$'' are for the other briber, as we care whether there are bribery operations for our briber to choose that would be winning for any bribery operations that the other briber might choose.
Moreover, the $\textrm{PossibleMove}$ predicates make sure that we only consider possible moves, the $\textrm{FeasibleMove}$ predicates make sure that we only consider feasible moves, and we make sure to update the current society by applying the $\textrm{ApplyMove}$ function.

\subsection{Finding Optimal Strategies}

Given the formula $\Phi$ as defined above and an initial society $\ves_1^1$, Theorem~\ref{theorem:pafpt} computes an initial move $\vem_1^1$ of a winning strategy of our briber if one exists, and otherwise reports that there is no winning strategy.
(In fact, the first predicate $\textrm{FeasibleMove}(\vem_1^1, B)$ may be removed and instead a more involved cost function may be minimized.)

To bound the complexity, we examine $L(\Phi)$, $\alpha(\Phi)$, and $\beta(\Phi)$.
The length $L(\Phi)$ is bounded by a polynomial in the number of variables, which is $\Oh(\ell \cdot \tau^2)$, since for each round we have a constant number of society and move vectors, which are of dimensions $\tau$ and $\tau^2$, respectively, and the number of rounds is $O(\ell)$.
The largest coefficient $\alpha(\Phi)$ is bounded by $\|\vecc\|_\infty$, which is the largest swap distance between two permutations, which is $\Oh(m^2)$.
It is crucial to note that the large input data, which is $B$ and $\ves_1^1$, only ever appear as constants (right hand sides), hence $\beta(\Phi) \leq \|B, \ves_1^1\|_\infty$.
Recall that Theorem~\ref{theorem:pafpt} implies an \FPT algorithm for parameters $\tau$ and $\ell$ if $L(\Phi)$ and $\alpha(\Phi)$ are bounded by a function of these, which indeed is the case here.

\section{Enriched Models}\label{section:enriched}

Above we described efficient algorithms in many cases, for the basic model described before.
This basic model is, however, quite restricted.
Luckily, our approach is very robust, thus rendering our algorithms quite general; indeed, as we show next, we have efficient algorithms in many cases for other, much less restricted models.

The basic observation is that, as we use PA, we can in fact design our formulas (in particular, $\Phi$), via a modular design; then, we can adapt each module separately to accommodate for various model generalizations and variants.
Let us first describe a modular and slightly more general version of $\Phi$:
\begin{align*}
  \Phi(\vem_1^1) \equiv&\ 
  \textrm{PreConditions}(\vem_1^1, \ves_1^1, B_1^1)
  \\
  &\land \ves_1^2 = \textrm{ApplyMove}(\vem_1^1, \ves_1^1) \\
  &\land (\bar{\ves}_1^2, B_2^1) = \textrm{PostProcessing}(\ves_1^2, \vem_1^1, B_1^1) \\
  \forall \vem_1^2 &:
  \textrm{PreConditions}(\vem_1^2, \bar{\ves}_1^2, B_1^2)
  \\
  &\land \ves_2^1 = \textrm{ApplyMove}(\vem_1^2, \bar{\ves}_1^2) \\
  &\land (\bar{\ves}_2^1, B_2^2) = \textrm{PostProcessing}(\ves_2^1, \vem_1^2, B_1^2) \\
  &\vdots \\
  \exists \vem_\ell^1 &:
  \textrm{PreConditions}(\vem_\ell^1, \bar{\ves}_\ell^1, B_\ell^1)
  \\
  &\land \ves_\ell^2 = \textrm{ApplyMove}(\vem_\ell^1, \bar{\ves}_\ell^1) \\
  &\land (\bar{\ves}_\ell^2, B_{\ell+1}^1) = \textrm{PostProcessing}(\ves_\ell^2, \vem_\ell^1, B_\ell^1) \\
  \forall \vem_\ell^2 &:
  \textrm{PreConditions}(\vem_\ell^2, \bar{\ves}_\ell^2, B^2_{\ell})
  \\
  &\land \ves' = \textrm{ApplyMove}(\vem_\ell^2, \bar{\ves}_\ell^2) \\
  &\land (\bar{\ves}', B_{\ell+1}^2) = \textrm{PostProcessing}(\ves', \vem_\ell^2, B^2_{\ell}) \\
  \textrm{Winn}&\textrm{ingConditions}(p, \bar{\ves}')   
\end{align*}

Specifically, for the basic model, $\textrm{PreConditions}$ merely checks both $\textrm{PossibleMove}$ and $\textrm{FeasibleMove}$, and the $\textrm{PostProcessing}$ step leaves the society and budgets intact. 
Below we discuss various enrichments to the basic model presented above and describe how to define $\textrm{PreConditions}$ and $\textrm{PostProcessing}$ to formulate them, thus to allow for our algorithmic approach to operate on them as well.

\subsection{Complex Budgeting Schemes}

In the basic model, each briber had a fixed budget to be used separately for each turn.
Some other options are discussed below.

\mypara{Initial Budgets}
We can allow the budget to be fixed at the beginning of the process. This would correspond to a campaign manager setting aside some amount to be used during the whole campaign.

This can be formulated in the $\textrm{PostProcessing}$ step, by decreasing from $B_{i-1}^j$ the amount just used, i.e., setting $B_i^j = B_{i-1}^j - \vecc \vem_{i-1}^j$.

\mypara{Individual Budgets}
We can also easily make it so that our briber has a different budget than the other briber, by simply plugging this info into each $B_i^j$, $i \in [\ell], j \in \{1,2\}$.

\mypara{Chunked Budgets}
We can allow the budget to be given in chunks. This would correspond to, say, a campaign manager assigning some amount to be used for each month (if one turn means one month).

This can be formulated in the $\textrm{PostProcessing}$ step, by decreasing from $B$ the amount just used, in addition to adding to $B$ some amount, i.e., if $C_i^j$ is the contribution to briber $j$ in round $i$, then we would add the constraint $B_i^j = B_{i-1}^j - \vecc \vem_{i-1}^j + C_i^j$ to the $\textrm{PostProcessing}$ formula.

\mypara{Adaptive Budgets}
We can even allow the budget to be adaptive; for example, that in each turn, the increase in the budget is a linear function that depends on the Borda score of the preferred candidate $p_i$ (alternatively, on $p$ for our briber and on the inverse Borda score of $p$ for the other briber), i.e., $B_i^j = B_{i-1}^j - \vecc \vem_{i-1}^j + S_p(\ves_{i-1}^j)$, where $S_p(\ves_{i-1}^j)$ is the Borda score of $p$ in the society $\ves_{i-1}^j$.
This would correspond to voters donating to the campaign as it runs depending on the intermediate poll results.

More generally, the budget for each briber is decided based on the society just before he plays.
The dependence might be arbitrary, as long as it can be formulated in PA.

\subsection{Voter Behavior}

We can model several scenarios w.r.t.\ how voters respond to bribery.
This is allowed by the fact that if we start with a society with a small number of types, the following refinements do not increase the number of types too much.
When we speak of different cost functions, these would be a part of the $\textrm{PreConditions}$ check, and they need to be encoded by a linear function with small coefficients in order for Theorem~\ref{theorem:pafpt} to give a fixed-parameter algorithm.

\mypara{Loyal Voters}
One possibility is that, whenever a voter is bribed for the first time, then she stays loyal and would never be bribed again.
This can be encoded by, for each type $t \in [\tau]$, introducing a new type $t'$ which has the same preference order but the cost of moving it is infinite for every type except for itself.
Thus, the resulting society has $2\tau$ types.

\mypara{Semiloyal Voters}
A more relaxed notion of loyalty would be that, each time a voter is bribed, her price for being bribed again goes up.
This is encoded by, for each type $t \in [\tau]$, introducing $2\ell+1$ new types $t_0, t_1,  \dots, t_{2\ell}$, where a voter is of type $t_j$ if their preference order corresponds to type $t$ and they have been bribed $j$ times in the process.
The cost of moving a voter of type $t_j$ to type $t'_{j+1}$ can be set as needed to model the increase in cost, and the cost of moving from $t_j$ to $t'_{j'}$ for any $j' \neq j+1$ is $\infty$, unless $t=t'$ and $j=j'$, where the cost is zero.
(Infinite costs are \emph{not} modeled using a large enough integer as usual in some contexts, but rather by upper bounding by zero the corresponding coordinate of a move vector.)

\mypara{Greedy Voters}
A different option is that, in each turn, each voter remembers the bribing offer from our briber as well as the bribing offer from the other briber, and simply goes with the higher offer. Note that for this to be formulated we shall add the possibility of offering more money to bribe a certain offer.
To model this, we need a list of all possible offers a voter can get, which is bounded by our assumption that the cost functions are bounded.
Then, for each original voter type $t \in [\tau]$ and for each possible offer $o \in \N$, we create a new voter type $t_o$.
A voter is of type $t_o$ if their preference order is $t$ and if their last accepted bribe offer was $o$.
The cost function of moving from $t_o$ to $t'_{o'}$ can be set as needed to model the observed behavior, and is $\infty$ for every $o' \leq o$ unless $t=t'$.
(To be precise, this models the behavior where a briber is aware that paying $o$ is not enough to move a voter; if we wanted to model that a briber decides to spend money but it has no effect, we could replace $\textrm{PreConditions}$ with a function altering the move and the budget, so that we would charge the briber for the move he attempted to make, but only execute the move accepted by the voters.
This would still give a fixed-parameter algorithm, but would lead to more clutter in the formula.)

\subsection{Complex Pay-off Functions and Winning Conditions} \label{section:socialchoicesettings}

In the basic model we considered the single-winner Borda voting rule as defining the winning condition, and we implicitly considered a 0/1 pay-off function, getting a $0$ if our candidate loses and getting a $1$ if he wins.
We can allow for more complex winning conditions and pay-off functions, corresponding also to more advanced social choice settings.
When discussing different pay-offs, we encode this still as a decision problem: is there a strategy (or what is the cheapest strategy) achieving a given pay-off.
Some options are discussed next.

\mypara{Other Single-Winner Voting Rules}
We can use any single-winner rule that is ILP-definable, which includes rules such as Borda and any other scoring protocol, Copeland$^\alpha$, STV, Bucklin, Kemeny, Dodgson, Young, and many others~\cite{faliszewski2018opinion,knop2018unifying}.
It is also possible to encode that our aim is to have a margin of victory (MoV; see, e.g.,~\cite{xia2012computing}) of at least some value (positive or negative), which expresses a desire to win an election at least with a certain margin (positive MoV) or to not lose an election by more than a certain value (negative MoV).
The specifics of how this may be encoded are highly dependent on a given voting rule, but taking Borda as an example, we would enforce that $\bigwedge_{c \in C \setminus \{p\}} S_c < S_p + M$ where $M$ is the desired margin of victory.

\mypara{One-Dimensional Single-Winner Elections}
Consider a setting in which candidates, as well as the bribers, are embedded on a real line. Then, given some voting rule, we might be satisfied not only with having $p$ win the eventual election, but also in an outcome in which the winner of the election is not too far from $p$. This corresponds naturally to single-peaked elections.

In fact, for each briber and its corresponding ideal point, the distance induces a ranking or an approval set over the alternatives. Observe that only this induced ranking might be encoded in the payoff function; thus, we can accommodate not only 1D elections, but any metric-based elections.

The way to encode this in $\textrm{WinningCondition}$ is to first compute the approval set $\textrm{AS} \subseteq C$ of all candidates within some desired distance from the ideal point, i.e., we wish to achieve that the winner in $\bar{\ves}'$ lies in $\textrm{AS}$.
Then, the winning condition would be a disjunction of formulas expressing that $c \in \textrm{AS}$ is a winner in $\bar{\ves}'$.
Since the number of candidates is bounded, this disjunction will be of bounded length.

\mypara{Multiwinner Elections}
Using a multiwinner election we can formulate, e.g., the winning condition that requires $p$ to be in the winning committee, which is similar to the previous case.
We can also define additive utility:
Imagine our briber having a utility $u(c)$ for each candidate $c$; then we can formulate a winning condition requiring some minimum value of $\sum_{c \in C} u(c)$.
To model this, assume we have a predicate $\textrm{Winner}(\ves, c)$ which is true if $c$ wins in society $\ves$; then the condition that the utility is at least $U$ is simply $\sum_{c \in C} \textrm{Winner}(\ves, c) \cdot u(c) \geq U$.  
(A small caveat is that we either require $\max_c u(c)$ to be bounded so that we do not introduce large coefficients, or we precompute which committees $C' \subseteq C$ have utility at least $U$ and then the winning condition is a disjunction of formulas expressing that exactly $C'$ is the winning committee, for each $C'$ with $u(C') \geq U$.
Such a disjunction is again of bounded size by the fact that the number of candidates is small.)
  
 As a special case of this approach, we can also consider that our briber likes some subset of the candidates and then formulate the winning condition requiring some lower bound on the number of liked candidates in the winning committee.

\subsection{Adversarial Constraints}

In the basic model (Section~\ref{section:basic}) we considered a worst-case assumption on the side of our briber in which the other bribers were only concerned with interfering with the goal of our briber.
It is also natural to assume that each of the other bribers $A_i$, $i \in [2, k]$, has its own preferred alternative $p_i$, and the goal of briber $A_i$ is to make $p_i$ the winner of the election.
This goal may be expressed (in broad terms) in $A_i$ never considering moves that harm them, which means that $A_1$ does not need to ensure they will be able to respond to such moves.
Note that this is significantly different from the collusion case considered before in that $A_1$ has more ``maneuvering space'' by not having to be overly pessimistic with respect to his opponent's moves.

Below we discuss some specific possibilities, which depend on the social choice setting (see Section~\ref{section:socialchoicesettings}).

Note that now we cannot simply reduce the setting with several other bribers (i.e., with $k > 2$) to the setting with only one briber (i.e., with $k = 2$); thus, our complexity also depends on $k$.

\mypara{Simple single winner elections}
The adversarial constraints might correspond to ``do not decrease the margin of victory of $p_i$''.
To model this in the case of Borda, the $\textrm{PreCondition}$ for player $A_i$ would be 
\[\min_{c \in C \setminus \{p_i\}} \{S_{p_i}(\ves+ \veDelta(\vem)) - S_c(\ves + \veDelta(\vem))\} \geq \min_{c \in C \setminus \{p_i\}} \{S_{p_i}(\ves) - S_c(\ves)\},\]
where $\ves$ is the current society, $\vem$ is the considered move, and $S_c(\ves)$ is the Borda score of candidate $c$ in society $\ves$.
The $\min$ function for two variables can be defined as follows: $z = \min\{x,y\} \equiv \left((z = x) \vee (z = y)  \right) \wedge \left((z \leq x) \wedge (z \leq y)\right)$, and the generalization to more variables is straight-forward.

\mypara{1D single winner elections}
The adversarial constraint might be to not increase the distance between the ideal point of the briber and the winner of the election.
To model this in the $\textrm{PreCondition}$, we need to compute, for each player $A_i$, their approval sets $\textrm{AS}_q^i$ for each $q \in [m]$ corresponding to the ranking induced by distance from their ideal point, i.e., $\{p_i\} = \textrm{AS}_1^i \subset \textrm{AS}_2^i \cdots \subset \textrm{AS}_m^i = C$.
With those in hand, it remains to encode a function giving the smallest $q \in [m]$ such that the approval set $\textrm{AS}_q^i$ is attained (i.e., the current winner belongs to it but not $\textrm{AS}_{q-1}^i$), and then we may only consider moves which do not increase $q$.
Since the number of these approval sets is bounded by the number of players $k$ and the number of candidates $\ell$, and the necessary functions can be encoded in PA, we again obtain a fixed-parameter algorithm with respect to parameters $\tau, m$ and $k$.

\mypara{Multiwinner elections}
A natural adversarial constraint might be to not decrease the ranking of $p$ in the ranking over the alternatives induced by the score given to each alternative by the voting rule $\calR$.
The encoding of this would again be specific for a given voting rule $\calR$, but can be done for all the aforementioned voting rules.

\subsubsection{Agent Far-sightedness}
The purpose of this section was in getting away from the overly pessimistic assumption that other agents are somehow willing to harm themselves.
However, perhaps \emph{not} accounting for any moves of opponents by which they harm themselves is overly naïve and optimistic, because there may exist moves that cause harm in the short term but are in fact beneficial in the long run.
To account for this, we introduce the notion of $z$-far-sightedness: an agent is \emph{$z$-far-sighted}, $z \in \N$, if they are only willing to perform moves which are guaranteed (no matter what strategy other agents choose) to benefit them after $z$ rounds of the campaigning game or by the time the game ends (i.e., by the time the election is evaluated), whichever comes first.

With the knowledge of the total number of rounds $\ell$, we can implement $z$-far-sightedness in our model as part of the \textrm{PreConditions} check.
We only sketch the approach here because giving a full description would be notationally cumbersome.
So, we construct a formula $\Phi_{i,z,\ell'}$ expressing the $z$-far-sightedness condition for agent $i$ in round $\ell'$; the structure of $\Phi_{i,z,\ell'}$ is analogous to $\Phi$ but only considers $\min\{z,\ell-\ell'\}$ rounds, and it is constructed from the perspective of the considered agent.

However, we need to account for the fact that other agents are also far-sighted.
This means that we have to construct $\Phi_{i,z,\ell'}$ recursively.
A possible issue is where the recursion terminates: this is where we need the foreknowledge that the game ends in $\ell$ rounds.
Specifically, in order to construct $\Phi_{i,z,\ell'}$, we need to have constructed $\Phi_{i',z,\ell''}$, where $i'$ is ``the next agent'' $i' = i+1 \mod k$, and $\ell''$ is either $\ell'$ if $i$ was not the last agent to play in round $\ell'$, and $\ell'+1$ otherwise.
Hence, the recursion depth in the construction of $\Phi_{i,z,\ell'}$ is at most $k \cdot \ell$, and the length of the resulting formula is bounded by $\tau, k$, and $\ell$, again yielding an \FPT algorithm when parameterized by $\tau, k$, and $\ell$.

\subsection{Diffusion Processes}

In the basic model the only way by which votes have changed is through bribery operations.
Very naturally, we can accommodate other ways to change votes, most notably diffusion processes, in which voters are assumed to reside on a social network and update their votes based on the votes of their neighbors.
Faliszewski et al.~\cite{faliszewski2018opinion} have defined a very general class of opinion diffusion processes, called ILP-definable processes, which do not consider each voter individually but group them by types (note that the types may be more refined than just by preference orders).
Any ILP-definable diffusion process can be encapsulated within the $\textrm{PostProcessing}$ step.
We need to specify a number of diffusion steps to happen after each bribery operation, which translates into the length of the resulting $\textrm{PostProcessing}$ formula.
Note that some ILP-definable diffusion processes have been shown to converge in a number of steps bounded by the number of voter types, hence if such process is considered, we may allow full convergence to happen after each bribery step.
Otherwise, a time-bound between bribery steps perhaps translates into aa small number of diffusion steps.

\section{Discussion and Outlook}

We developed an algorithmic framework for situations in which several bribers aim at rigging a given election, and showed that in many cases, finding optimal strategies can be done in time which depends super-polynomially only on the number of alternatives in the election or, more generally, only on the number of voter types in the given election.
Our framework is versatile and incorporates many features which make the resulting model highly realistic, in particular we can efficiently solve situations in which there are \begin{enumerate*}[label=\bfseries(\arabic*)] \item complex voter types; \item different budgets, including adaptive budgets, for each agent; \item voters are embedded in a network and are affected by certain diffusion processes; \item various realistic restrictions apply to the manipulating agents.\end{enumerate*}
Next we discuss some avenues for future research.

\mypara{Unbounded number of bribery rounds}
Our results are parameterized by the number of rounds of the bribery-counterbribery exchange.
It is known that deciding Presburger arithmetic sentences (a problem easier than \textsc{Presburger arithmetic minimization}) has super-exponential complexity unless the length of the formula is bounded~\cite{FR}, and even with bounded length but large coefficients, the complexity grows increasingly higher in the polynomial hierarchy with increasing number of quantifier alternations~\cite{nguyen2019short}.
However, it remains open whether any of the proposed campaigning models are also hard when the number of rounds is not bounded.
We conjecture that the most basic model with unit-cost swap actions and identical budgets for each round is \FPT for a constant number of bribers and parameterized by the number of candidates, because we believe that the identical budget will enforce some sort of repetitive structure in the strategies of bribing agents.
However, we also conjecture that already when we allow the budgets to very from round to round, the problem is \NP-hard for constantly many bribers and candidates.

\mypara{Presburger arithmetic and AI}
Finally,
Theorem~\ref{theorem:pafpt} which we rely on, and particularly Cooper's algorithm for Presburger arithmetic, is a general and widely applicable tool.
We are curious to see more applications in computational social choice, and, more generally, in the field of Artificial Intelligence.
This is particularly interesting because Theorem~\ref{theorem:pafpt} falls into the larger context of identifying fixed-parameter tractability in problems above NP in the polynomial hierarchy (of which PA is an example) which has only recently begun to be explored, and which has great relevance to AI~\cite{de2016parameterized}.

\begin{acks}
	Koutecký was partially supported by Charles University project UNCE/SCI/004 and by the project 19-27871X of GA ČR.
\end{acks}

\bibliographystyle{named}
\bibliography{bib}

\end{document}